\documentclass[aps,prb,twocolumn,superscriptaddress]{revtex4-1}
\usepackage[colorlinks=true,citecolor=blue,linkcolor=blue,urlcolor=blue]{hyperref}
\usepackage{graphicx, nicefrac, textcomp}
\usepackage{hyperref}

\newcommand{\CRO}{Ca$_2$RuO$_4$}

\begin{document}

\title{Raman scattering from current-stabilized nonequilibrium phases in Ca$_2$RuO$_4$}

\author{K. F{\"u}rsich}
\author{J. Bertinshaw}
\author{P. Butler}
\author{M. Krautloher}
\author{M. Minola}
\email{M.Minola@fkf.mpg.de}
\author{B. Keimer}
\email{B.Keimer@fkf.mpg.de}
\affiliation{Max-Planck-Institut f{\"u}r Festk{\"o}rperforschung, Heisenbergstrasse 1, 70569 Stuttgart, Germany}

\date{\today}

\begin{abstract}
We used Raman light scattering to study the current-stabilized nonequilibrium semimetallic and metallic phases in \CRO. By determining the local temperature through careful analysis of the Stokes and anti-Stokes intensities, we find that Joule heating can be completely avoided by supplying sufficient cooling power in a helium-flow cryostat, and that the current induces the semimetallic state without inducing any significant heating. We further investigate the current-induced semimetallic state as a function of temperature and current. We confirm the absence of long-range antiferromagnetic order and identify a substantial Fano broadening of several phonons, which suggests coupling to charge and orbital fluctuations. Our results demonstrate that the semimetallic state is a genuine effect of the applied electrical current and that the current-induced phases have characteristics distinct from the equilibrium ones.
\end{abstract}

\pacs{df}

\maketitle

Controlling and stabilizing electronic phases in correlated-electron materials are pivotal objectives of modern solid-state science \cite{Hwang12, Basov17}. In this context, compounds with $4d$ valence electrons have attracted a lot of attention as a plethora of ground states can be observed \cite{Nakatsuji00}. In particular, the fine balance between electron correlations, spin-orbit coupling and structural distortions in $4d$ metal oxides manifests itself as a high susceptibility to external perturbations. One prominent example is the layered perovskite system (Ca$_{1-x}$Sr$_x$)$_2$RuO$_4$ whose end members, \CRO$\;$and Sr$_2$RuO$_4$, are a Mott insulator and an unconventional superconductor, respectively \cite{Maeno94, Carlo12}.

\CRO$\;$shows a temperature-driven insulator-to-metal transition (IMT) at $T_{\text{IMT}}=357\,\text{K}$ \cite{Braden98, Alexander99, Friedt01}. The concomitant structural phase transition preserves the lattice symmetry (space group \textit{Pbca}), but yields elongated RuO$_6$ octahedra with a reduced tilt and an enlarged $c$ lattice parameter in the high-temperature metallic state. The insulating and metallic phases are therefore named \textit{``S''-Pbca} and \textit{``L''-Pbca} phase, respectively, where $S$ and $L$ refer to short and long $c$ axes. In addition to temperature, other parameters, such as Sr substitution \cite{Carlo12}, epitaxial strain \cite{Dietl18} and hydrostatic pressure\cite{Nakamura02}, can turn \CRO$\;$metallic.

The key role of octahedral rotation, compression and tilts is further exemplified in the magnetic state of \CRO, which sets in at $T_{\text{N}}=110\,\text{K}$ and is characterized by an A-type antiferromagnetic (AFM) ordering \cite{Braden98}. Several theoretical\cite{Khaliullin13, Akbari14, Chaloupka16} and experimental \cite{Jain17, Souliou17} studies point towards an unconventional mechanism for magnetic ordering in \CRO, resulting from the condensation of spin-orbit excitons. The ``excitonic magnetism" model further predicts that \CRO$\;$lies close to a quantum critical point separating antiferromagnetic and spin-orbit singlet states, which was confirmed by the experimental observation of a soft longitudinal amplitude mode of the spin-orbit condensate.

Recent research has revealed new perspectives for \textit{in-situ} control of nonequilibrium phases in \CRO$\;$by electrical currents. In their pioneering work Nakamura \textit{et al.}\cite{Nakamura13} found a current-induced IMT at electrical fields much smaller than those required for a conventional Zener breakdown. X-ray diffraction measurements indicated that the current-induced metallic phase, also called $L^*$, is different from the temperature-driven metallic $L$-phase. Further studies focused on the current-induced closing of the Mott-gap \cite{Okazaki13} and discussed the nonequilibrium phases of \CRO$\;$in close relation to nonequilibrium superconductors and charge-ordered insulators. Interestingly, a stripe pattern separating metallic and insulating regions was recently observed by scanning near-field optical microscopy with the stripe orientation perpendicular to the direction of the applied current \cite{Zhang19}. This unusual phase separation was found at intermediate current densities, \textit{i.e.} smaller than the critical value to induce the IMT. These results were explained in terms of long-range strain effects resulting from the strong coupling between electronic and structural degrees of freedom in \CRO. Importantly, all experiments including the spatially resolved ones suggest that the current-induced metallic state is a nonfilamentary bulk phenomenon.

Besides the IMT observed at currents of about $100\,\text{mA}$, Sow \textit{at al.} detected a large diamagnetic response \cite{Sow17} at much smaller current densities and below $T=50\,\text{K}$. The authors suggest a phenomenological model in which the gap closing by electrical current gives rise to a semimetallic state with small electron and hole pockets. A detailed study based on neutron crystallography in combination with ab-initio calculations indeed found a partially gapped Fermi surface, thereby indicating a semimetallic electronic structure in the current-induced state \cite{Bertinshaw18}. 

However, the driving mechanism behind the closing of the Mott gap by DC currents, which is considered the essential ingredient for the strong diamagnetism, is yet to be clarified together with the possible role played by Joule heating. This has motivated the spectroscopic study we report here. The current-stabilized steady state opens the rare possibility to perform inelastic photon scattering experiments in a nonequilibrium state that would be extremely challenging in light-induced transient phases \cite{Lee18}. In this article, we use Raman light scattering to probe the magnetic, electronic and structural properties of the current-stabilized nonequilibrium phases of \CRO. Additionally, we exploit the unique capability of Stokes and anti-Stokes Raman scattering to probe the temperature of the sample \textit{in-situ}. We rule out Joule heating as the origin of the current-driven IMT in actively cooled environments and demonstrate that the current-stabilized phases are distinct from the equilibrium ones.

\begin{figure}
\includegraphics[width=0.999\columnwidth]{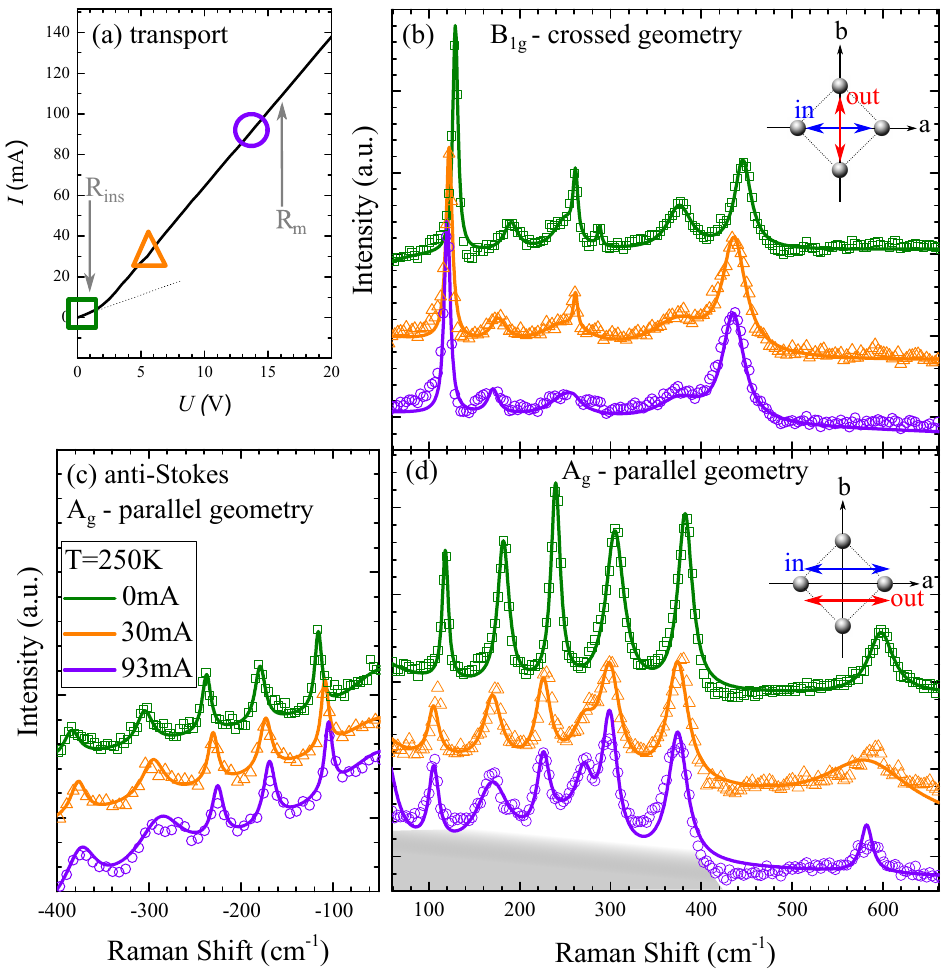}
\caption{Current-induced IMT in \CRO. (a) Transport measurement. $R_{\text{ins}}$ ($R_{\text{m}}$) indicates regions with high (low) resistance, which corresponds to the insulating (metallic) phase. The dotted line is an extrapolation of the $R_{\text{ins}}$ region. (b) Raman spectra in $B_{1g}$ geometry. (c) and (d) Raman spectra in $A_{g}$ geometry for anti-Stokes and Stokes parts of the spectrum, respectively. The symbols in panel (b),(c) and (d) represent experimental Raman data, taken at the current values indicated with the corresponding marker in panel (a). The lines in panels (b),(c) and (d) are fits to experimental data. The inset in panels (b) and (d) illustrate the scattering geometry of the Raman experiment with respect to the Ru lattice (dotted lines: Ru-O-Ru bonds; $a$, $b$: crystallographic axes). The shaded area in panel (d) illustrates the increased electronic response, as discussed in the text.}
\label{S-AS}
\end{figure}

High-quality \CRO$\;$single crystals with $T_{\text{N}}=110\,\text{K}$ were grown by the floating zone method as previously described \cite{Jain17, Nakatsuji01}. The \CRO$\;$crystals with dimensions $2.6\times1.0\times0.6\,\text{mm}^3$ were mounted on the sample holder of a He flow type cryostat using GE varnish \footnote{See Supplemental Material at [URL will be inserted by publisher]}. 
The Raman measurements were performed with a Jobin-Yvon LabRam HR800 single-grating Raman spectrometer using the $632.8\,\text{nm}$ excitation line of a HeNe laser. 
All spectra were taken in backscattering geometry along the crystallographic $c$ axis, while the current was applied in-plane along the crystallographic (110) direction. As \CRO$\;$crystallizes in the orthorhombic $Pbca-D^{15}_{2h}$ space group, excitations in the $B_{1g}$ and $A_g$ representations of the point group $D_{2h}$ were probed in crossed and parallel configurations, corresponding to $z(XY)\bar{z}$ and $z(XX)\bar{z}$ geometries in Porto's notation, respectively. The polarization of the incident light was thus always kept $45\,^\circ$ away from the Ru-O-Ru in-plane bonds [see inset in Fig. \ref{S-AS} (b) and (d) for the scattering geometry]. Some spectra were corrected for the Bose thermal factor to obtain the Raman response $\chi''(\omega)$. Our spectra measured in the absence of current flow are in agreement with the ones reported in literature \cite{Rho05, Souliou17}

\begin{figure}
\includegraphics[width=0.999\columnwidth]{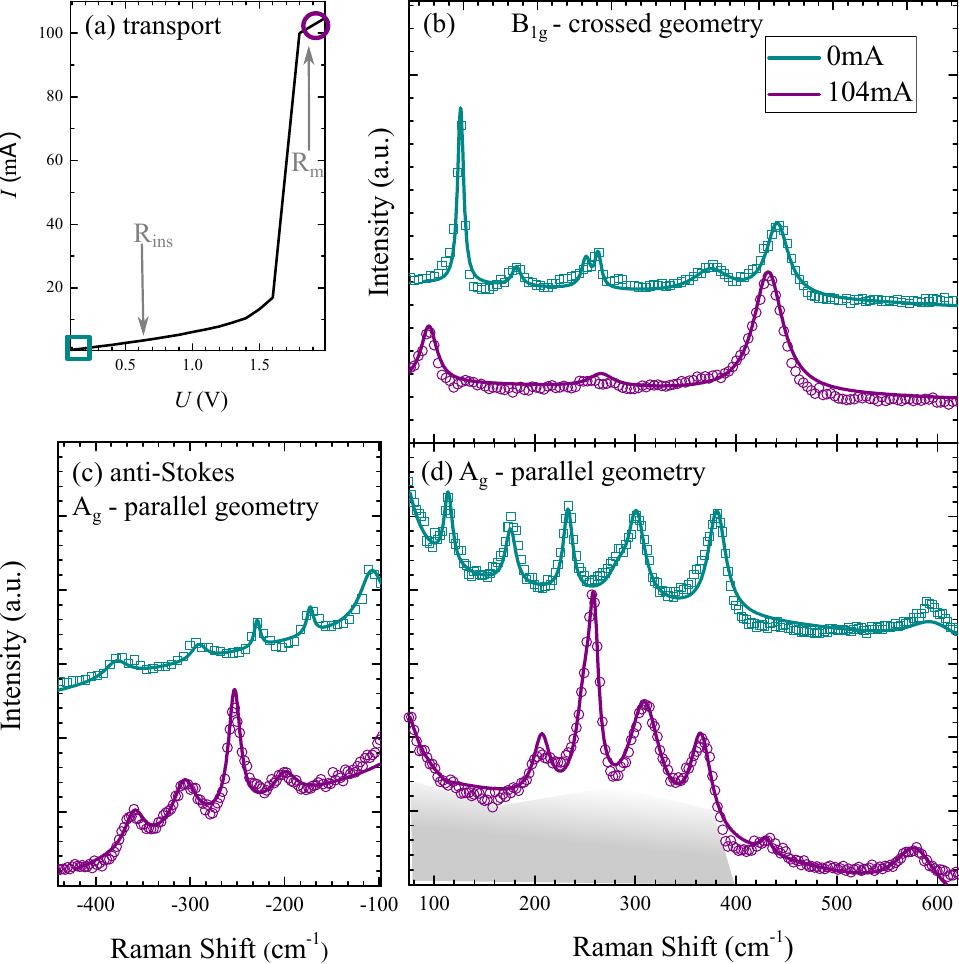}
\caption{Temperature driven IMT in \CRO. (a) Transport measurement. (b) Raman spectra in $B_{1g}$ geometry. (c) and (d) Raman spectra in $A_{g}$ geometry for anti-Stokes and Stokes parts of the spectrum, respectively. Temperatures are given in Tab. \ref{T_summary}. In analogy to Fig. \ref{S-AS}, the current values for the Raman spectra (experimental data: symbols; fit: lines) are indicated with the corresponding symbol in panel (a). In panel (d) the shaded area illustrates the increased electronic response.}
\label{g_heating}
\end{figure}

In the first part of this article, we focus on the current-induced IMT at and near room temperature, where the metallic phase sets in at currents of about $100\,\text{mA}$  (i.e. in the high-current regime). In particular, we choose two configurations to evaluate the role of Joule heating. First we stabilize the temperature at $T=250\,\text{K}$ and provide sufficient cooling power to avoid the Joule heating effects discussed below (Fig. \ref{S-AS}). In a second set of experiments, we do not provide any cooling power to the sample and we leave it at room temperature, $T=295\,\text{K}$, before applying the electrical current (Fig. \ref{g_heating}). As the voltage across the sample is increased \cite{Note1}, we observe a nonlinear increase in currents for both experimental conditions [Figs. \ref{S-AS}(a) and \ref{g_heating}(a)]. Raman spectra were measured at different points in the transport curve and show distinct changes for lower resistance values due to the appearance of the metallic phase [see Figs. \ref{S-AS} and \ref{g_heating}]. Particularly, the increased electronic response towards lower wavenumbers [shaded gray area in Figs. \ref{S-AS}(d) and \ref{g_heating}(d)] indicates the transition into the metallic state \cite{Ponosov12, Devereaux07}.

We first focus on the careful evaluation of the temperature in the current-induced metallic state by comparing Stokes and anti-Stokes intensities \cite{Cardona, Hayes12, Herman11}, which correspond to creation and annihilation of an elementary excitation and therefore have a positive or negative Raman shift, respectively. The ratio of the Raman scattered photons of both processes can be written as \cite{Cardona, Hayes12}:
\begin{equation}
\frac{I_{\text{AS}}}{I_{\text{S}}}=\frac{(\nu_{\text{L}}+\nu_{\text{p}})^3}{(\nu_{\text{L}}-\nu_{\text{p}})^3} \exp({\frac{-h\nu_{\text{p}}}{k_{\text{B}}T}})
\label{eq_SAS}
\end{equation}
with the frequencies of the phonon mode $\nu_{\text{p}}$ and of the laser $\nu_{\text{L}}$, the Planck constant $h$ and the Boltzmann constant $k_{\text{B}}$. Based on this relation, one can determine the \textit{in-situ} temperature $T$ at the laser spot.

\begin{figure}
\includegraphics[width=0.8\columnwidth]{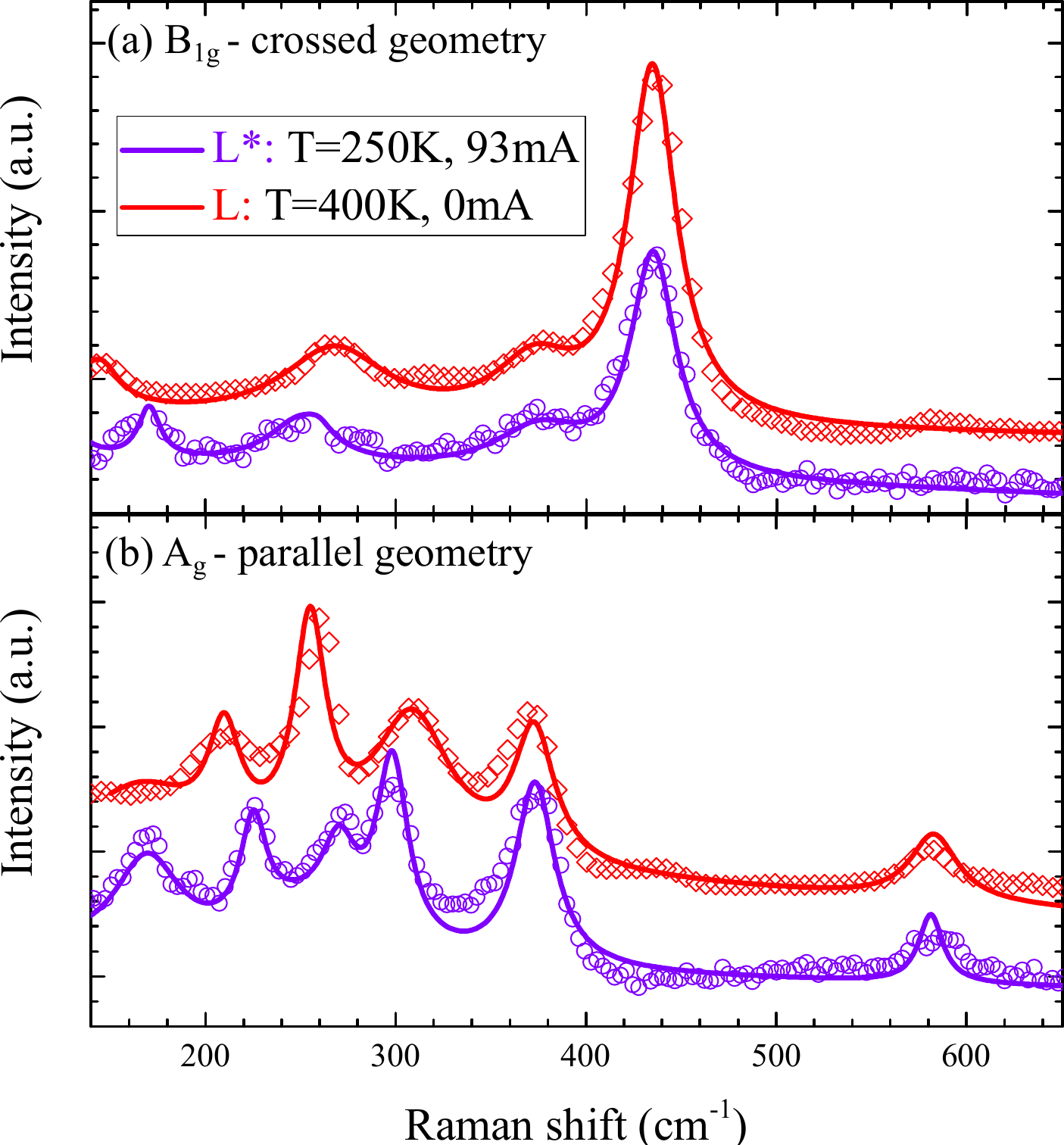}
\caption{Raman scattering from metallic phases in \CRO. (a), (b) Raman spectra in $B_{1g}$ and $A_{g}$ geometry, respectively. Symbols (Lines) represent experimental (fitted) data.}
\label{LvsL*}
\end{figure}

Following equation \ref{eq_SAS}, we extract the temperatures for all phonon frequencies $\nu_{\text{p}}$\cite{Note1}. The averaged values of the \textit{in-situ} determined temperatures are given in Tab. \ref{T_summary}. Our analysis shows that with active cooling (Fig. \ref{S-AS}), the current-induced IMT is a true effect of the electrical current, while heating can be neglected (left part of Tab. \ref{T_summary}) . Remarkably, the measurements at room temperature without cooling (Fig. \ref{g_heating}) reveal a significant Joule heating (right part of Tab. \ref{T_summary}) to a temperature $T>T_{\text{IMT}}$. Without applying cooling power in fact the Stokes anti-Stokes analysis shows that the IMT is not induced by electrical- current but rather by Joule heating \footnote{The Raman spectra in fact are also closely similar to those collected in the equilibrium $L$-phase rather than to those from the current-induced $L^*$ phase shown in  Fig. \ref{LvsL*}.}. 
This observation is in agreement with a recent study by G. Mattoni \textit{et al.} using a thermal imaging setup \cite{Mattoni19}. As a consequence, in order to observe the nonequilibrium metallic $L^*$-phase one must provide sufficient cooling power.

\begin{table}[h]
\caption{Temperatures determined following equation \ref{eq_SAS} as \CRO$\,$transitions from the insulating to the metallic phase.}
\begin{ruledtabular}
\begin{tabular}{l c c | l c c }
  \multicolumn{3}{c |}{with cooling}			  & \multicolumn{3}{c}{no cooling} \\
  \colrule
$I$ / mA	&  $T$ / K  & phase  & $I$ / mA	&  $T$ / K & phase  \\
\colrule
0		&  $249\pm4$			& $S$ 			& 0		&  $295\pm4$			& $S$ \\
30		&  $249\pm5$			& transition\\
93		&  $252\pm5$			& $L^*$		&104		&  $390\pm5$			& $L$ \\
\end{tabular} 
\end{ruledtabular}
\label{T_summary}
\end{table}

Having answered the fundamental question about the presence and role of heating effects, we now turn to the comparison of nonequilibrium and equilibrium metallic phases. Fig. \ref{LvsL*} compares the $L^*$-phase to the metallic equilibrium $L$-phase obtained by increasing the temperature to $T=400\,\text{K}$, well beyond $T_{\text{IMT}}=357\,\text{K}$\cite{Note1}. The spectra of the $L$- and $L^*$- phase are similar (Fig. \ref{LvsL*}), and both are clearly distinct from the ones of the $S$-phase [Figs. \ref{S-AS} and \ref{g_heating} (b),(d) for $0\,\text{mA}$]. It is worth mentioning that the Raman spectra of the $L^*$- phase resemble those of metallic Sr$_2$RuO$_4$ measured at $T=300\,\text{K}$ \cite{Sakita01}. However as observed in Fig. \ref{LvsL*} (and in agreement with the neutron scattering study of Ref. \onlinecite{Bertinshaw18}, where the sample was mounted in the same way), we note that $L$- and $L^*$- phases are not identical. Differences in the number and positions of the phonon modes are particularly apparent in the $A_g$ channel [Fig. \ref{LvsL*}(b)]. This could be due to differences in the crystal structure of the two phases, which were inferred from structural refinements of neutron diffraction measurements for both nonequilibrium and equilibrium metallic states \cite{Bertinshaw18, Friedt01}. The $L^*$-phase shows a larger orthorhombicity (defined as the difference between the in-plane lattice parameters $a$ and $b$) and a smaller out-of-plane lattice parameter $c$ in comparison to the $L$-phase. Both features suggest that the $L$-phase is less distorted than the $L^*$-phase and hence, the Raman spectra for these states could reflect this difference. In light of recent results \cite{Zhang19}, another possibility to explain the differences between $L$- and $L^*$-phases revolves around the heterogeneous development of the current-stabilized state. In this scenario, the nonequilibrium metallic phase does not pervade the bulk of the sample in a uniform way, thereby resulting in defects, phase-boundaries and phase-slips. Due to phonon-band folding induced by periodic domain boundaries, additional phonon modes might appear in the Raman spectrum. Future spatially-resolved experiments might be able to discriminate between these possibilities.

\begin{figure}
\includegraphics[width=0.999\columnwidth]{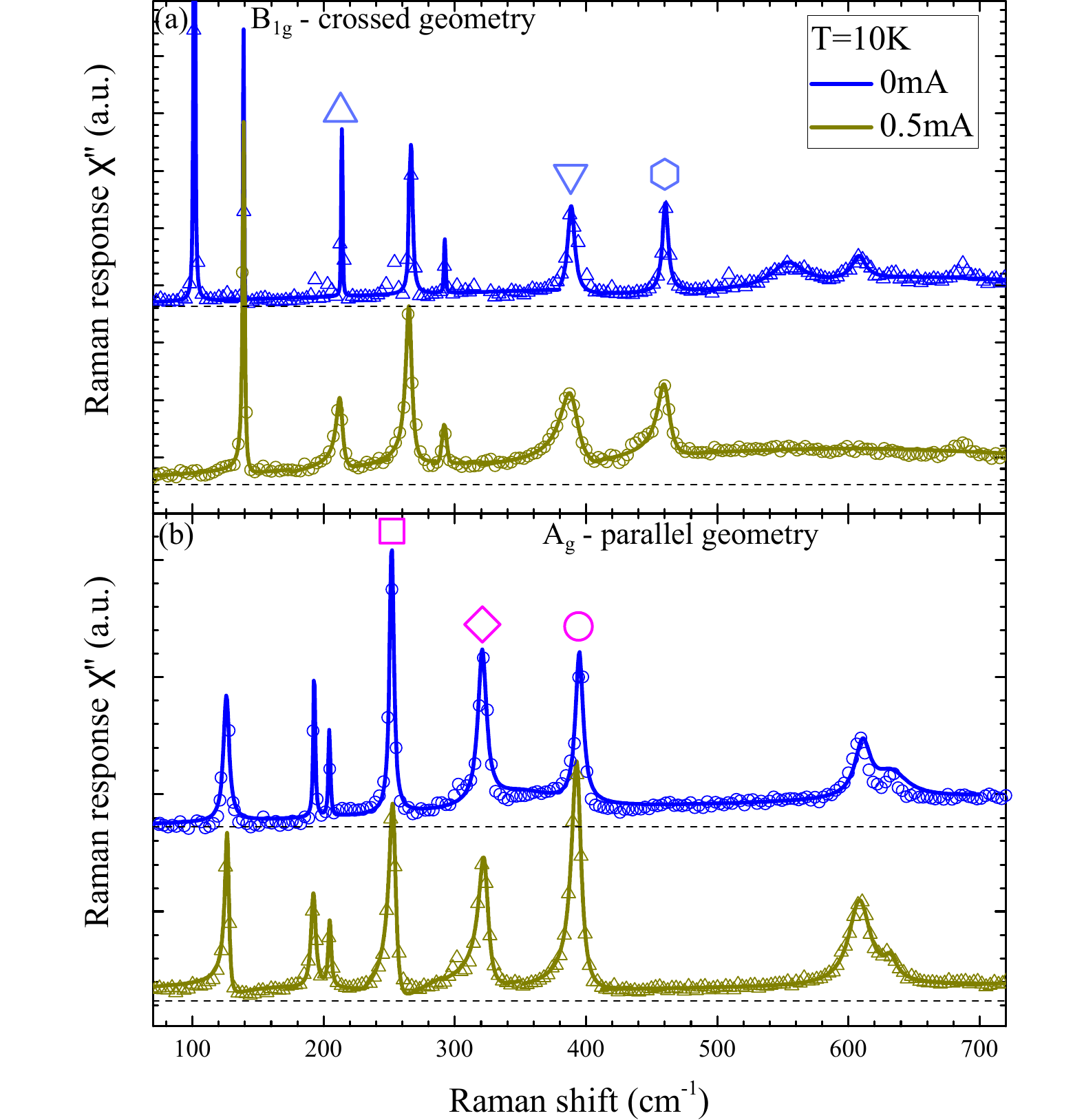}
\caption{Current-stabilized semimetallic diamagnetic phase in comparison to the equilibrium AFM insulating state. Raman spectra in (a) crossed $B_{1g}$ geometry and (b) $A_{g}$ geometry. The lines are superpositions of Fano peaks fitted to the data (small symbols). The phonon modes marked with large symbols are analyzed in Fig. \ref{Fano}.}
\label{SvsS*}
\end{figure}
We now turn to the discussion of the low-current regime which is characterized by currents smaller than $1\,\text{mA}$ and is distinct from the high current states discussed until now.  As mentioned before, the low currents lead to the formation of the so called $S^*$-phase at low temperatures. We compare the nonequilibrium $S^*$-phase to the equilibrium $S$-phase at $T=10\,\text{K}$ \cite{Note1}. The Raman spectra for crossed and parallel configurations are shown in Fig. \ref{SvsS*}. 
The feature at $101\,\text{cm}^{-1}$ in $B_{1g}$ geometry was previously identified as single-magnon excitation\cite{Souliou17} inherent to the soft-moment magnetic state in \CRO. In the current-stabilized phase the magnon peak is absent, thereby suggesting that long-range AFM order is suppressed. This observation confirms earlier bulk magnetometry measurements \cite{Sow17}, and neutron studies \cite{Bertinshaw18} which did not identify magnetic Bragg reflections characteristic of AFM ordering in the $S^*$-phase.

The phonon energies are nearly identical in the $S$- and  $S^*$-phases (apart from some minor softening in the  $S^*$ phase), which is consistent with the nearly identical Ru-O bond lengths identified by neutron diffraction \cite{Bertinshaw18}.

\begin{figure}
\includegraphics[width=0.999\columnwidth]{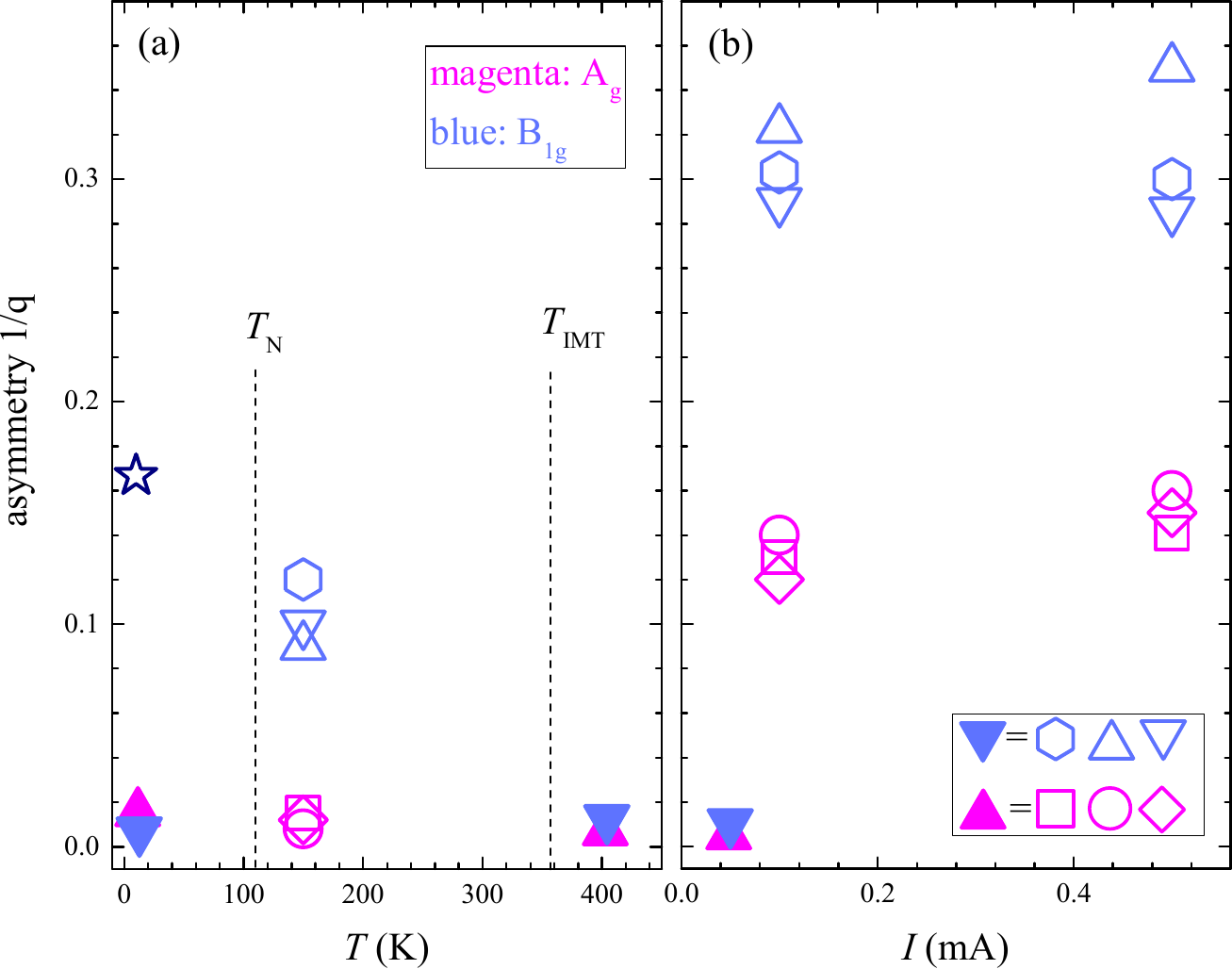}
\caption{Asymmetry parameter $1/q$ for (a) equilibrium and (b) nonequilibrium phases of \CRO. Data for panel (b) were taken at $T=10\,\text{K}$. Magenta (blue) symbols correspond to $1/q$ values for parallel (crossed) polarization. For the $B_{1g}$ geometry, we analyze the phonons at 211, 388 and $458\,\text{cm}^{-1}$ and for the $A_{g}$ channel  at 252, 321 and $392\,\text{cm}^{-1}$, as indicated in Fig. \ref{SvsS*} with the corresponding open symbol. Filled symbols summarize all phonons for a specific geometry. The increased asymmetry in the insulating paramagnetic state is in qualitative agreement with Ref. \onlinecite{Rho05}. The dark blue star marks the asymmetry value for Sr-doped \CRO$\;$at $456\,\text{cm}^{-1}$ ($B_{1g}$ channel), taken from Ref. \onlinecite{Rho03}.}
\label{Fano}
\end{figure}
In contrast to the symmetric phonon lineshapes in the $S$-phase, several of the phonons in the $S^*$-phase exhibit markedly asymmetric lineshapes. Phonon asymmetries are known to result from interactions with a continuum. The quantitative evaluation of this interaction calls for a detailed fitting analysis and comparison to the equilibrium phases. The phonon asymmetry can be obtained from fits to a Fano profile: $I(\nu)=I_0(q+\epsilon)^2/(1+\epsilon)^2$, with $\epsilon=(\nu-\nu_{\text{p}})/\Gamma$. Besides the phonon frequency $\nu_{\text{p}}$  and the effective phonon linewidth  $\Gamma$, we extract the Fano parameter $q$ whose inverse is directly proportional to the electron-phonon coupling strength and the imaginary part of the electronic susceptibility \cite{Naler02, Rho05}. In Fig. \ref{Fano}, we compare the asymmetry parameter $1/q$ of selected phonons (as indicated in Fig. \ref{SvsS*})  for $S^*$- and equilibrium phases of \CRO. Interestingly, we find an increased asymmetry for the current-stabilized phase suggesting an enhanced electronic susceptibility to orbital and/or charge fluctuations \cite{Mizokawa01, Sakaki13}. For low and high temperatures in the equilibrium phases and for low currents in the nonequilibrium phase, we find nearly negligible asymmetry parameters for all phonons investigated.

The presence of charge and orbital fluctuations has already been discussed in the equilibrium phases of \CRO. The Fano asymmetry in the paramagnetic insulating state of \CRO$\;$was attributed to orbital fluctuations\cite{Rho05, Zegkinoglou05}, similar to observations in Ir- and Ti- based correlated oxides \cite{Gretarsson16, Ulrich15}. Sr-doped metallic \CRO$\;$exhibits a larger asymmetry
 due to coupling to charge fluctuations \cite{Rho03, Sakita01}. Our data show that the asymmetry $1/q$ in the nonequilibrium $S^*$-phase is even larger, possibly because both orbital and charge channels contribute in the semimetallic state. Additionally, we point out that the ratio of asymmetry parameters for $B_{1g}$ and $A_g$  is different in the nonequilibrium and equilibrium phases. In the current-stabilized states the  $1/q$ is larger for the $A_g$ channel, which further supports the appearance of charge fluctuations. Detailed theoretical work is required to single out and quantify the different contributions from charge and orbital fluctuations.

In summary, we studied the nonequilibrium phases of \CRO$\;$using Raman light scattering. Comparing Stokes and anti-Stokes Raman signals demonstrates directly that the nonequilibrium phases can be reached via a true current-driven transition, discarding Joule heating effects in the presence of sufficient cooling power. In addition, our data have revealed pronounced Fano asymmetries of several phonons and the current-driven disappearance of the magnon excitations in the nonequilibrium $S^*$- phase. These observations are signatures of an electronic structure that is different from any of the previously studied equilibrium phases.

In the future, it will be interesting to extend our Raman experiments to nonzero momentum transfer by using intermediate-energy resonant inelastic x-ray scattering at the Ru $L$ edges \cite{Suzuki19}. It would be also beneficial to perform scattering experiments with applied electrical current on ruthenates and other compounds that are less insulating in nature, such as the Ti-substituted bilayer ruthenate Ca$_3$(Ru$_{1-x}$Ti$_x$)$_2$O$_7$ \cite{Sow19}, as this would allow to explore more in detail the low current regimes.

\begin{acknowledgments}
We thank I. Mazin, Y. Maeno, G. Mattoni, J. Porras and H. Suzuki  for fruitful discussions, and A. Schulz for technical assistance. We acknowledge support from the European Research Council under Advanced Grant No. 669550 (Com4Com) and from the Deutsche Forschungsgemeinschaft (DFG, German Research Foundation) - Projektnummer 107745057 - TRR 80.
\end{acknowledgments}

\bibliographystyle{apsrev4-1}
\bibliography{Literature}

\end{document}